\documentclass[%
 aps,
 floatfix,
 amsmath,amssymb,
 reprint,%
 longbibliography
]{revtex4-1}

\usepackage{graphicx}
\usepackage{dcolumn}
\usepackage{bm}
\usepackage{textgreek}
\usepackage{multirow}
\usepackage[colorlinks=true,allcolors=blue,breaklinks=true]{hyperref}%
\usepackage[hyphenbreaks]{breakurl}
\usepackage{todonotes}
\usepackage{soul}
\usepackage{tabularx}

\usepackage[labelformat = empty,position=top]{subcaption}
\usepackage[export]{adjustbox}

\newcommand{\mysubfigimg}[3][,]{%
  \setbox1=\hbox{\includegraphics[#1]{#3}}
  \leavevmode\rlap{\usebox1}
  \rlap{\hspace*{30pt}\raisebox{\dimexpr\ht1-3\baselineskip}{#2}}
  \phantom{\usebox1}
}

\begin{document}

\preprint{AIP/123-QED}

\title{Discriminating Uranium Isotopes Based on\\ Fission Signatures Induced by Delayed Neutrons}

\author{K.~Ogren}
\email{kogren@umich.edu}

\author{J.~Nattress}

\thanks{Currently at Oak Ridge National Laboratory, Oak Ridge, TN 37831 USA}

\author{I.~Jovanovic}
\email{ijov@umich.edu}

\affiliation{ 
Department of Nuclear Engineering and Radiological Sciences, University of Michigan, Ann Arbor, MI 48109 USA
}%

\date{\today}

\begin{abstract}
The use of active interrogation to induce delayed neutron emission is a well-established technique for the characterization of special nuclear materials (SNM). Delayed neutrons have isotope-characteristic spectral and temporal signatures, which provide the basis for isotope identification. However, in bulk materials that contain an appreciable fissile (\textit{e.g.,} $^{235}$U or $^{233}$U) fraction, such as highly-enriched uranium (HEU), delayed neutrons have a high probability of inducing additional fissions. As a result, the overall delayed neutron signature consists of two distinct components: the ``primary'' delayed neutrons (emitted directly by fission fragments), and the ``secondary prompt'' fission neutrons produced in fission induced by primary delayed neutrons. These prompt products differ from ``primary'' delayed neutrons both in their energy spectra and in the presence of coincident radiation released by the parent fission event. The presence and relative quantity of prompt products from delayed fission depend on the cross-section of the material in the energy range of delayed neutrons, which may differ significantly between isotopes, thus providing an exploitable means for isotope differentiation. In this work, we demonstrate two experimental approaches for discriminating between $^{235}$U and $^{238}$U isotopes based on the measurement of delayed neutron-induced fission products. First, HEU and depleted uranium objects are differentiated through the detection of high-energy prompt neutrons from delayed fission using both recoil-based organic liquid scintillators and thermalization spectra from a custom-built capture-gated composite detector. Secondly, coincident radiation measurements are used as the basis for discrimination by comparing the overall rates and time evolution of fission events when delayed neutrons are present. 
\end{abstract}


\maketitle
\section{Introduction}
Measurement methods that can provide detailed information on the composition of special nuclear materials (SNM) are integral to many nuclear security and nonproliferation applications. In particular, determination of the relative isotopic abundance of $^{235}$U and $^{238}$U plays a central role in international safeguards inspections, the production and accounting of nuclear fuel, and verification of proper storage and dismantlement of weapons components under disarmament treaties~\cite{PANDA,Berndt2010,Runkle2010}. Active interrogation sources use external radiation to induce nuclear reactions in a target material, and are commonly employed to characterize the content of SNM by inducing fission to augment both the prompt and delayed neutron and gamma-ray signatures emitted by the material~\cite{Runkle2012}.

When SNM is interrogated using a pulsed neutron source, the induced fission signature persists even after the source is turned off. Source neutrons become thermalized in the surrounding material, and their population decays with a characteristic time on the order of a few microseconds~\cite{Beckurts1957}. When SNM is present, these thermalized neutrons can induce additional fission events, providing a secondary source of neutrons, which causes the overall population of thermalized neutrons to decay more slowly. Measurements of the neutron dieaway can thus provide an indication of the presence of fissile material, and are a well-established means for detecting and characterizing SNM~\cite{Caldwell1982,Jordan2007}. Unlike the common dieaway technique, this work focuses on another persistent signal: the emission and subsequent interaction of delayed neutrons in the interrogated material. By recording the sample response over a much longer timescale (tens of seconds), we show that it is also possible to deduce additional signatures from $^{235}$U and $^{238}$U and gain insight into the isotopic content of the sample.

The emission of delayed neutrons by fission fragments is a well-known phenomenon~\cite{Roberts1939further,Roberts1939delayed,Bohr1939dropmodel}, and the measurement of delayed neutron signatures has long been established as an effective method for detecting fissionable materials~\cite{Amiel1962}. Fission reactions generate a variety of unstable nuclei, which are typically neutron-rich and undergo $\beta$ decay. A fraction of the decaying nuclei also release their energy through neutron emission. The delay between the initial fission event and these secondary emitted neutrons depends on the chain decay kinetics, which are characteristic of the fission fragments, and ranges from a few hundreds of nanoseconds to tens of seconds. The $\beta$-delayed neutrons are commonly divided into a set of groups based primarily on similarities in their precursor half-lives~\cite{Keepin1957}. The specific parameters for the groups depend on the isotope undergoing fission, as well as the type and energy of the fission-inducing particle. Despite the fact that delayed neutrons account for only a small fraction of the overall fission neutron yield, detection systems that utilize active interrogation to intensify the emitted delayed neutron signal have been successfully used to detect and identify fissionable materials~\cite{Keepin1969,Jones2003,Slaughter2003,Jovanovic2018,Chichester2009,Mayer2016}.

Because the delayed neutron groups for a particular fissionable isotope have a unique set of individual decay constants, each isotope possesses a characteristic aggregate temporal profile for delayed neutron emission, which can be used as the basis for identification. Previous studies have applied this principle to differentiate SNM samples by measuring the decay of the delayed neutron rate for short-lived~\cite{Kinlaw2006} and long-lived groups~\cite{Li2004,Myers2006,Sellers2012}. In our own prior work, we have utilized both the buildup and decay time profiles of long-lived delayed neutron groups to perform isotopic discrimination and infer the enrichment level of uranium~\cite{Nattress2018}. While the energy and timing characteristics of delayed neutrons have been determined by dedicated precision measurements, the observed delayed neutron signature for bulk materials can be complicated by additional interactions before the delayed neutrons escape the object. In the case of fissile materials such as $^{235}$U and $^{239}$Pu, delayed neutrons with an average energy of 250--450~keV can readily induce numerous additional fission events. This delayed neutron-induced fission is a basic concept in nuclear reactor kinetics, where it represents an important consideration in maintaining the desired state of criticality in a reactor. In the context of neutron energy spectrum, however, it leads to an overall delayed signal that is a superposition of two components: ``primary'' delayed neutrons, which are emitted directly from the decay of fission fragments, and the prompt fission neutrons, where the fission is induced by delayed neutrons. In fact, in order to measure the delayed neutron energy spectra with a high degree of accuracy, sample sizes have been restricted to small amounts ($<$10~g) of material with the express purpose of limiting distortions caused by fission multiplication~\cite{Keepin1957,Shalev1973,Batchelor1956}. 

Because the prompt fission products of the delayed neutron-induced fission events are emitted nearly instantaneously, they mimic the time distribution of the delayed neutrons, and do not significantly alter the measured neutron temporal profile (notwithstanding the potential differences in detector efficiency when measuring the time-evolving neutron spectrum). In contrast, the overall delayed energy spectrum is significantly changed by the introduction of prompt fission neutrons, which typically have much higher energies than delayed neutrons. Furthermore, the prompt neutrons from delayed fission are accompanied by additional coincident neutrons and $\gamma$ rays. The determining factor in the relative abundance of prompt fission products in the delayed signal is the average fission cross-section of a particular material in the delayed neutron energy range. Because this cross-section may differ significantly between isotopes, as is the case for $^{235}$U and $^{238}$U, the detection of high-energy prompt neutrons and coincident radiation from fission in the delayed signature of SNM may provide the basis for isotopic identification.

In the previous work it has been demonstrated that proton beam are an effective means for inducing delayed neutron signatures in SNM targets and can discriminate fissionable materials at lower dose rates than photon- or neutron-based sources~\cite{Morris2010}. The report also proposes that measurement of delayed neutrons above a certain energy threshold would indicate the presence of delayed neutron-induced fission and provide a method for discriminating uranium isotopes. 

Here, we present two experimental methods for disambiguating the genesis of delayed neutrons as a means for discriminating $^{235}$U and $^{238}$U. In the first approach, which leverages the method proposed in Ref.~\cite{Morris2010}, recoil-based organic liquid scintillators and a custom-built capture-gated composite detector are used to perform spectroscopic measurements of delayed neutrons, and highly-enriched uranium (HEU) is successfully differentiated from depleted uranium (DU) based on the presence of high-energy prompt neutrons in its delayed signature. While the measurement of high-energy prompt neutrons from delayed neutron-induced fission has previously been proposed as a method for discriminating uranium isotopes, to the best of our knowledge, this work represents the first time that such energy information has been specifically targeted and extracted from the overall delayed signature as a means for isotopic identification. In the second approach, we demonstrate the first use of coincidence counting to observe the contribution of delayed neutron-induced fission to the overall fission rate and successfully differentiate HEU from DU on this basis.

\section{Materials \& Methods}

Experimental measurements were performed at the Device Assembly Facility (DAF), Nevada National Security Site, using highly-enriched uranium (HEU) and depleted uranium (DU) test objects of approximately 13.8~kg and 12.8~kg, respectively. Further details on the uranium objects are included in the Supplementary Information. Each uranium object was interrogated with 14.1-MeV neutrons produced by a Thermo Scientific P211 DT neutron generator, with an approximate isotropic yield of 10$^8$~n/s. The objects were placed at a distance of 13~cm from the generator, as measured from the center of the object to the center of the target plane in the generator tube. The DT generator was operated at a pulse rate of 100~Hz with a pulse width of approximately 10~\textmu s, which was consistent across all measurements.

In each measurement, the uranium object was surrounded by an array of detectors, which included one 5.1-cm diameter NaI(Tl) detector, two 7.6-cm diameter Eljen EJ309 organic liquid scintillators~\cite{EJ309}, and one custom-built heterogeneous composite scintillator. The composite detector is a larger version of the prototype described in Ref.~\cite{Mayer2015}, and consists of an array of lithium glass square rods embedded in a cylindrical matrix of scintillating polyvinyl toluene (PVT). The principal detection mechanism for the composite detector is neutron capture by $^6$Li in the glass rods, which possess very different scintillation properties from the PVT plastic. As a result, neutron capture events are easily distinguishable by both a characteristic pulse shape and the characteristic $Q$-value of the reaction. The PVT matrix surrounding the lithium glass rods serves a dual purpose. Not only does it increase the capture efficiency of the detector by moderating the incident neutrons, but the scintillation response of the PVT to proton recoils in the neutron thermalization process provides a signal whose magnitude is correlated to the incident neutron energy~\cite{Shi2016}. By exploiting the time coincidence between a capture pulse and the preceding proton recoil pulse, spectroscopic neutron energy analysis can be performed~\cite{Nattress2016}. Additional information on the composite detector design and operation is provided in the Supplementary Information. 

\begin{figure}[ht]
\centering
\includegraphics[width=\linewidth]{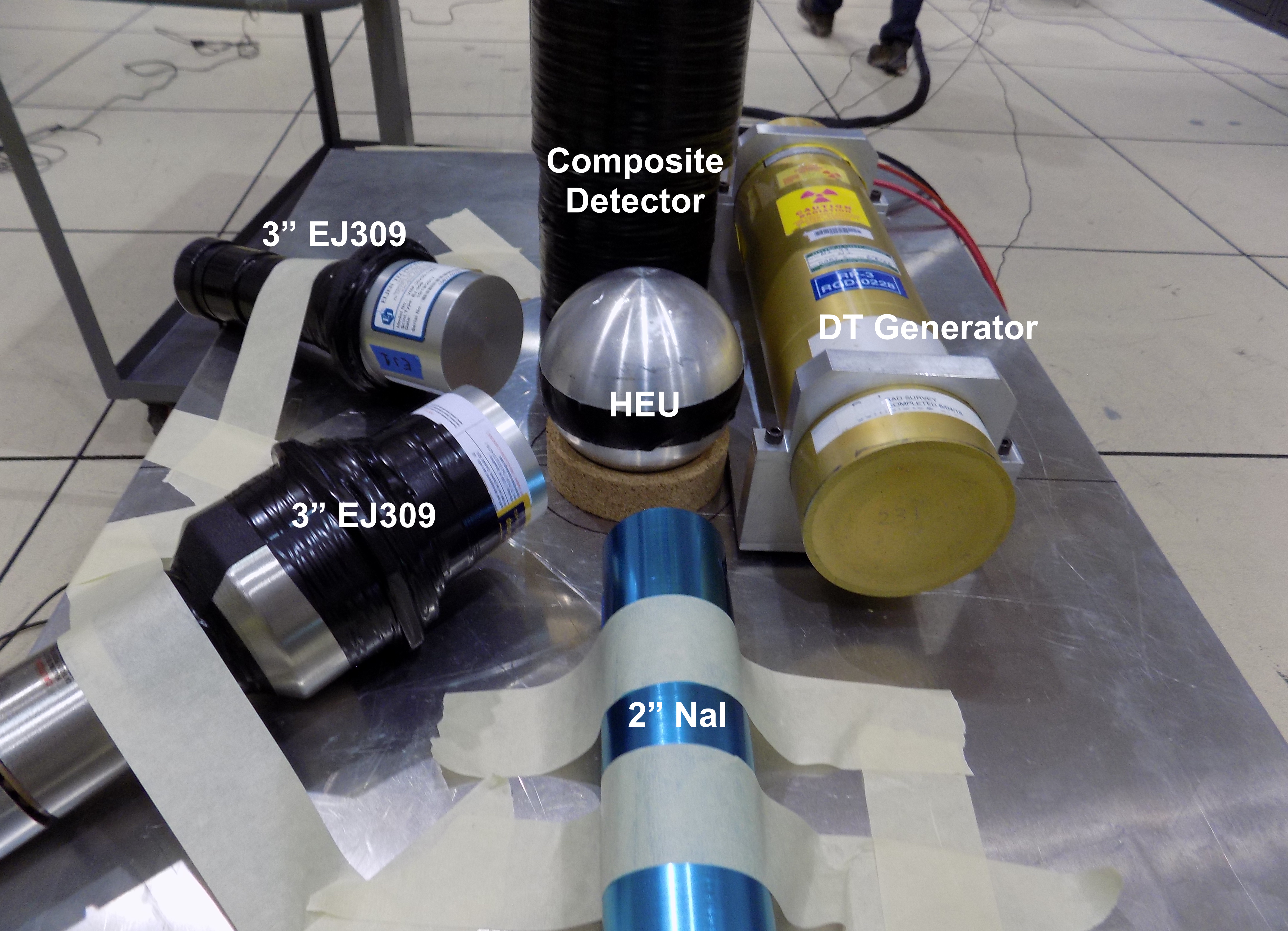}\hspace{1cm}
\caption{Experimental setup during measurement of the HEU object.}
\label{fig:Setup}
\end{figure}

For each measurement, the composite detector was placed vertically at a distance of 21~cm from the central axis of the PVT cylinder to the center of the uranium object. The NaI(Tl) and EJ309 scintillators were placed at different locations around the object, each at a distance of 11~cm from the center of the object to the front face of the detector. Fig.~\ref{fig:Setup} shows the experimental setup used for measurement of both uranium objects.

Digital pulses were recorded using CAEN desktop waveform digitizers, and data acquisition and storage was performed using CAEN Multi-Parameter Spectroscopy Software (CoMPASS)~\cite{COMPASS}. For each waveform, short-gate ($Q_{\mathrm{short}}$) and long-gate ($Q_{\mathrm{long}}$) charge integrals were recorded to provide the basis for pulse shape discrimination. The integration parameters were optimized for each detector prior to the experiment to provide the best degree of discrimination. Further details on the data acquisition equipment and methods are presented in the Supplementary Information.

The neutron generator was operated in a series of on/off cycles, during which the induced delayed neutron signatures of the HEU and DU objects were recorded. In each cycle, the generator was turned on for one minute, then off for one minute. Each object was interrogated over a period of approximately 2.5~hours ($\sim$70 on/off cycles), and the data collected during the generator off time was aggregated to form the overall delayed signal. Passive measurements of the HEU and DU objects were also recorded, for 3 minutes and 10 minutes, respectively. The detectors were calibrated using $^{137}$Cs and AmBe sources.

\section{Simulation}

To estimate the expected contrast in the delayed neutron emission spectra from bulk samples of HEU and DU, Monte Carlo simulations were conducted. The simulation and the associated method for separating the components of the delayed neutron spectrum are detailed in the Supplementary Information. For HEU, prompt neutrons from delayed fission account for about 65\% of the delayed neutron signature, with the remaining 35\% contributed by primary delayed neutrons. In contrast, primary delayed neutrons made up over 98\% of the emitted signal for DU.

\begin{figure}[ht]
\centering
\includegraphics[width=\linewidth]{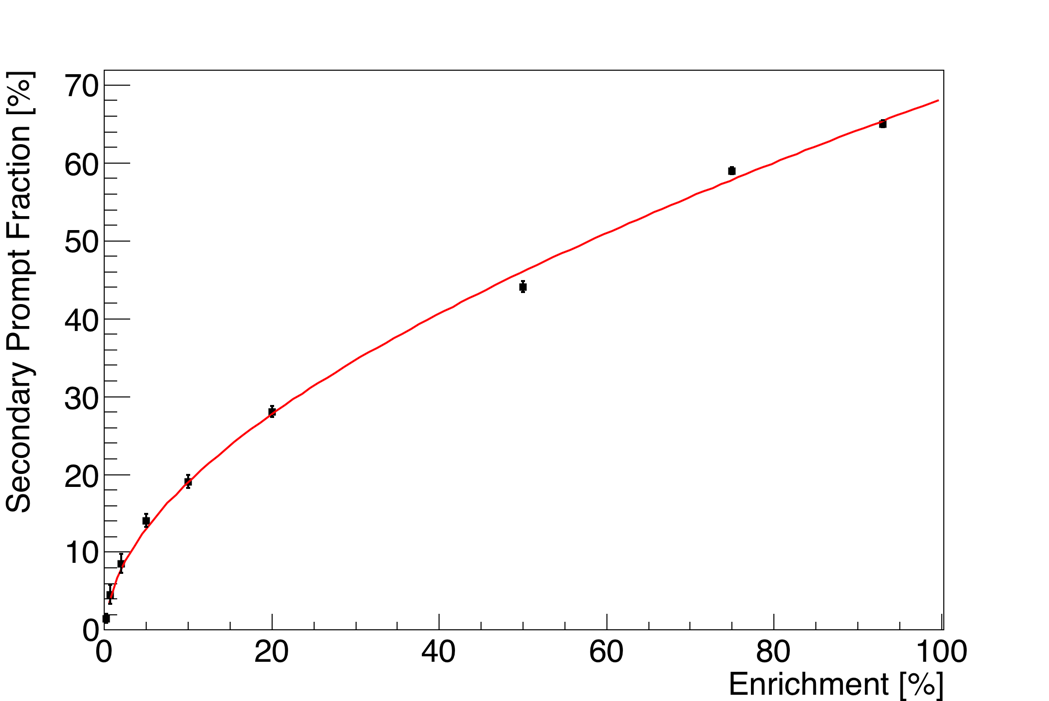}\hspace{1cm}
\caption{Simulated contribution of secondary prompt neutrons to the overall delayed neutron signature as a function of uranium enrichment. The red line is provided only as a guide to the eye.}
\label{fig:SecondaryFrac}
\end{figure}

In addition to the experimental object materials, simulations of the expected relative proportion of primary delayed neutrons and prompt neutrons from delayed fission were performed for a variety of uranium enrichments to explore the potential for finer determination of the enrichment level based on the induced neutron spectrum. Fig.~\ref{fig:SecondaryFrac} shows the proportion of the overall delayed neutron signal that is expected to be contributed by secondary prompt neutrons as a function of enrichment. Table~\ref{table:NeutronSpecPercents} shows a summary of the simulated neutron spectrum breakdown for selected enrichment levels. 

\begin{table}[ht]
\parbox{1\linewidth}{
\caption{Simulated contribution to overall delayed neutron spectrum from primary delayed neutrons and prompt neutrons from delayed fission for selected uranium enrichments.}
\centering
\begin{tabular}{ccc}
\hline
\hline
Material (Enrichment) & Primary Delayed & Prompt \\
\hline
DU (0.02\%) & \ \ 98.5\% \ \  &  1.5\%  \ \  \\
LEU (5\%) & \ \  86\%\ \  &  14\% \ \   \\
LEU (20\%) & \ \ 72\% \ \  &  28\% \ \   \\
HEU (93\%) & \ \ 35\% \ \  &  65\% \ \   \\
\hline
\hline
\end{tabular}
\label{table:NeutronSpecPercents}
}
\end{table}

While higher-energy prompt neutrons make up only a small fraction of the delayed neutron spectrum for DU, their proportion increases rapidly as enrichment increases from 0.2\% to 20\%, and they already form a significant part of the delayed signal for 5\%-enriched LEU. As a result, it may be possible to distinguish LEU from natural uranium based on the presence of higher-energy neutrons in the delayed neutron signature. Furthermore, the relative contribution from prompt neutrons doubles as enrichment increases from 5\% to 20\%, and more than doubles again for weapons-grade enrichment levels ($>$90\%). Such separation suggests that the proportion of higher-energy delayed neutrons may serve as an observable for estimating the enrichment level of uranium-containing materials.

\begin{figure}[!h]
  \centering
  \begin{tabular}{@{}p{1.0\linewidth}@{}p{1.0\linewidth}@{}}
    \mysubfigimg[width=\linewidth]{\hspace{4.5cm}\color{black}(a)}{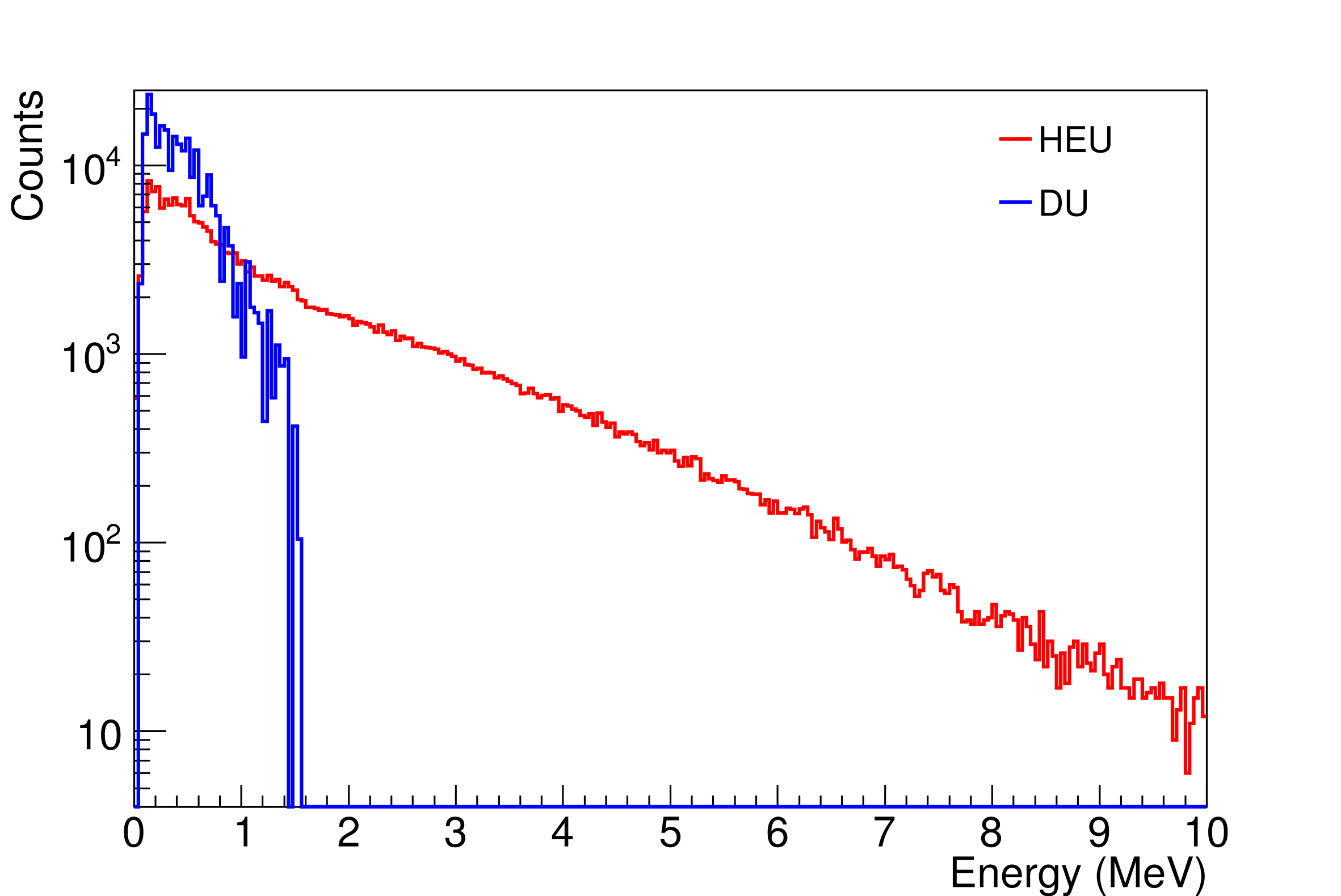} \\
    \mysubfigimg[width=\linewidth]{\hspace{4.5cm}\color{black}(b)}{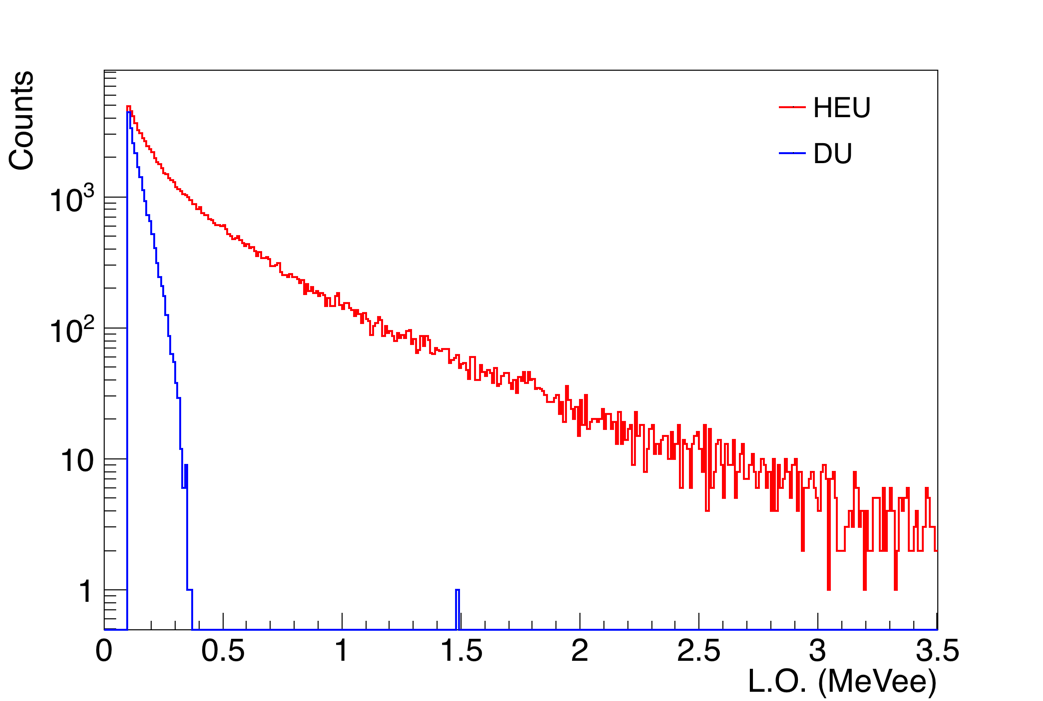}  \\
  \end{tabular}
  \caption{(a)~Simulated delayed neutron energy spectra for bulk HEU and DU, based on proportional contribution of prompt products of delayed fission, and (b)~simulated response of the composite detector to overall HEU and DU delayed energy spectra. In both cases, counting results are based on simulation of 250,000 source particles.}
  \label{fig:DN_spec_Sims}
\end{figure}

For each material, the fractional contributions of primary delayed neutrons and prompt neutrons from delayed fission were used to approximate the delayed neutron energy spectra, which were then used to simulate the expected response in the composite detector using the Geant4 framework~\cite{Agostinelli2003}. While MCNPX is better suited for simulating the production of delayed neutrons, Geant4 does not have the same geometrical limitations, making it the more convenient choice for modeling the complex structure of the composite detector. To simulate the light output response to proton recoils during thermalization in the PVT, the detector was bombarded with neutrons with energies sampled from the delayed~\cite{Keepin1957} and prompt~\cite{Cranberg1956} energy spectra in accordance with their relative proportion for each isotope. The light output produced by neutron elastic scatters on protons in the detector was modeled using a similar method to the one described in Ref.~\cite{Enqvist2013}, with a polynomial function of the form
\begin{equation}
    L = aE - b\left[1-\textrm{exp}(-cE)\right],
    \label{eq:LO}
\end{equation}
where $L$ is the light output, $E$ is the energy deposited on the proton, and $a$, $b$, and $c$ are fitting parameters. The light output contribution from scatters on carbon nuclei was assumed to be approximately 2\% of the energy deposited. A Gaussian broadening function was parameterized and applied to the calculated light output according to the method outlined in Ref.~\cite{Nattress2017cal}. Fig.~\ref{fig:DN_spec_Sims} shows the simulated delayed neutron energy spectra for bulk HEU and DU, as well as the expected light output response of the composite detector. The light output units are MeVee (MeV electron equivalent), where 1~MeVee represents the light output generated by 1~MeV of electron energy deposition.

The simulated delayed energy spectra and corresponding detector response show very significant differences for each material based on the presence of delayed fission events, especially at higher neutron energies. This suggests that $^{235}$U and $^{238}$U should be readily distinguishable based on the presence of high-energy neutrons in the measured delayed signal.

\section{Experimental Results \& Discussion}

\begin{figure}[!ht]
\centering
\includegraphics[width=\linewidth]{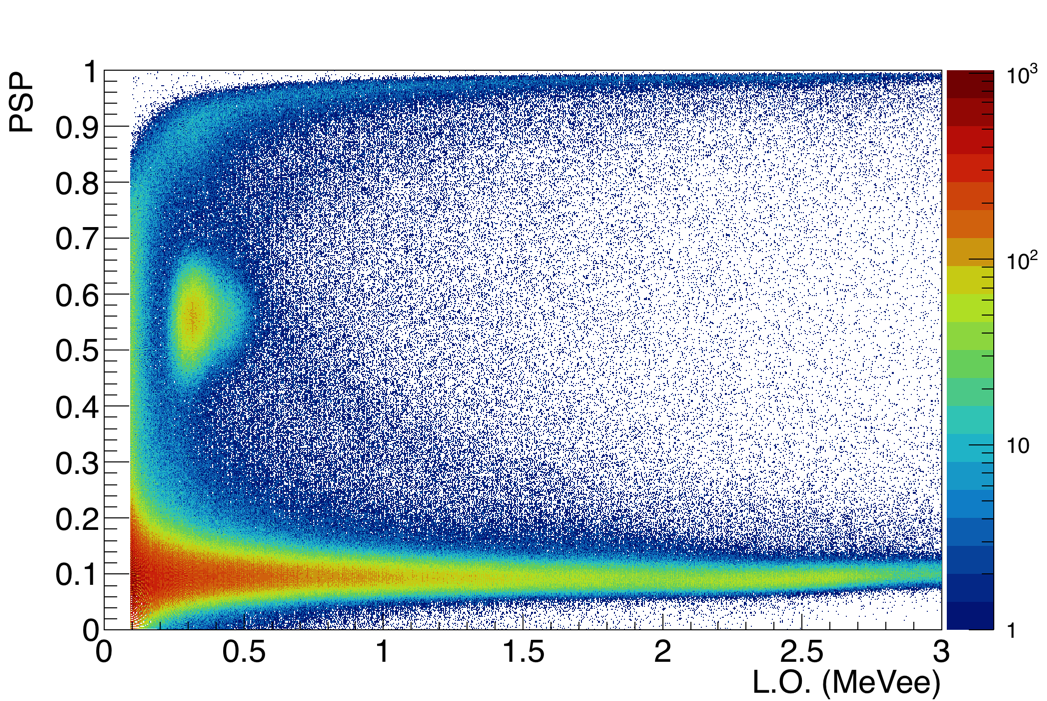}\hspace{1cm}
\caption{Calibration $PSP$ and light output distribution in the composite detector when exposed to an AmBe source. Neutron captures were identified using a 3-$\sigma$ cut around the island feature centered at $PSP=0.55$ and light output of 0.32~MeVee.}
\label{fig:PSPcal_ambe}
\end{figure}

A significant advantage of the composite detector is that it provides strong discrimination of neutron-capture events on $^6$Li, which can then be used to identify potential preceding thermalization events in the detector and extract spectroscopic energy information from the incident neutrons. Fig.~\ref{fig:PSPcal_ambe} shows the pulse-shape parameter ($PSP$) and light output distribution for the AmBe calibration measurement in the composite detector, where the $PSP$ is defined as
\begin{equation}
PSP = \left(Q_{\mathrm{long}} - Q_{\mathrm{short}}\right)/Q_{\mathrm{long}}.
\label{eq:EQ_PSP}
\end{equation}
The parameter space located around $PSP=0.55$ and light output of about 0.32~MeVee corresponds to neutron capture events. A 3-$\sigma$ cut was established in two dimensions around this region, and any events falling within the cut were classified as neutron captures in the subsequent measurements.

\begin{figure}[!ht]
\centering
\includegraphics[width=\linewidth]{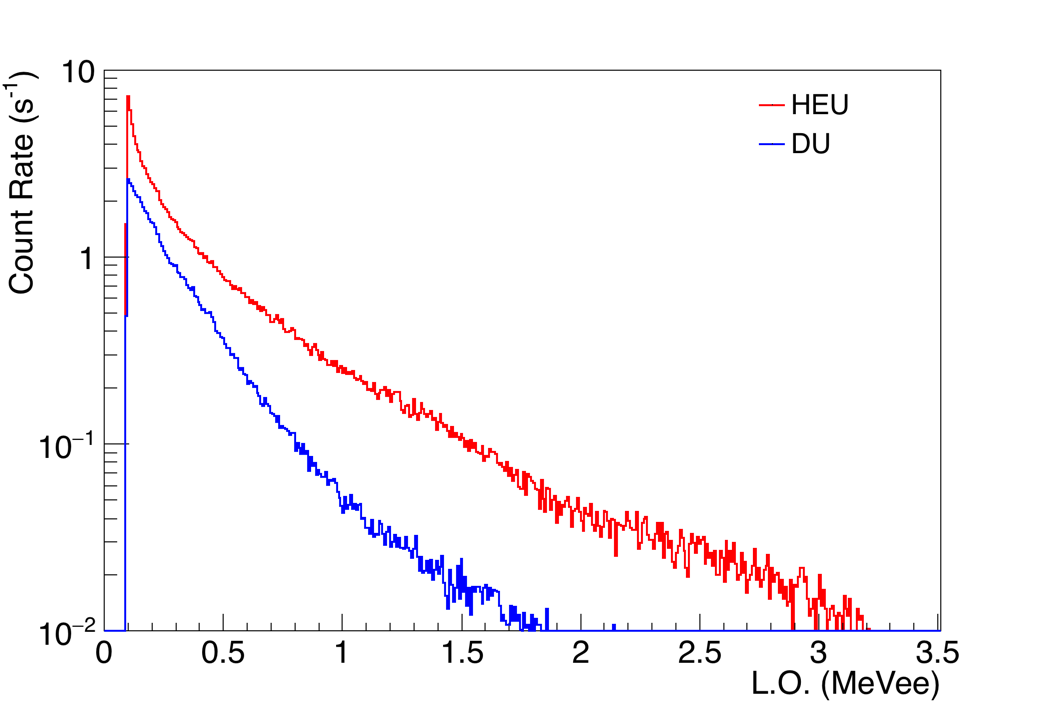}\hspace{1cm}
\caption{Comparison of experimentally measured capture-gated light output distributions in the composite detector for HEU and DU. Recoil events with $PSP = 0.0865 \pm 3\sigma$ and a recoil-capture coincidence time $<76$~\textmu s were accepted. The count rate for DU above 400~keVee is consistent with the measured rate of gamma-ray accidentals within the recoil-capture acceptance window.}
\label{fig:thermHEUvsDU}
\end{figure}

The delayed neutron energy spectra for HEU and DU were compared by analyzing the capture-gated light output response in the composite detector. For each neutron capture event, the previously recorded pulse was examined to determine if it could have been caused by thermalization of the fast neutron in the PVT prior to capture. Because the type of PVT used in the composite detector is not PSD-capable, there is only one recoil region corresponding to both neutron and gamma-ray interactions. A Gaussian fit to this region established a mean $PSP$ value of 0.0865, and pulses exhibiting deviation from the mean greater than 3-$\sigma$ were rejected. Geant4 simulations were also used to determine the time scale of neutron thermalization in the composite detector. Both prompt and delayed incident neutron energy spectra were modeled; the results indicate that 99\% of captures occur within 76~\textmu s of the initial scattering interaction in the detector, and that incident neutron energy has little effect on the shape of the time distribution of capture-gated recoil pulses. As such, only recoil pulses that occur within 76~\textmu s before the subsequent capture event were included in the capture-gated light output distribution. Fig.~\ref{fig:thermHEUvsDU} shows the resulting capture-gated light output distributions for HEU and DU.

While some increase in the delayed neutron counting rate is expected for HEU relative to DU due to the greater mass of the HEU object and increased fission cross section for 14.1-MeV neutrons, the higher overall rate is also consistent with increased multiplication caused by delayed neutrons. However, the marked increase in high-light-output events for HEU indicates a significant difference in the overall energy spectrum, which can be explained by the presence of higher-energy prompt neutrons from delayed fission. Such differences in the shape of the neutron energy spectrum may provide the basis for discrimination between isotopes and inference of the enrichment level. One possible method would involve comparing the relative contribution of high-light-output events to the overall spectrum using a simple ratio of two integration regions. A more detailed outline of the proposed method is provided in the Supplementary Information.

\begin{figure}[!ht]
  \centering
  \begin{tabular}{@{}p{1.0\linewidth}@{}p{1.0\linewidth}@{}}
    \mysubfigimg[width=\linewidth]{\hspace{0.5cm}(a)}{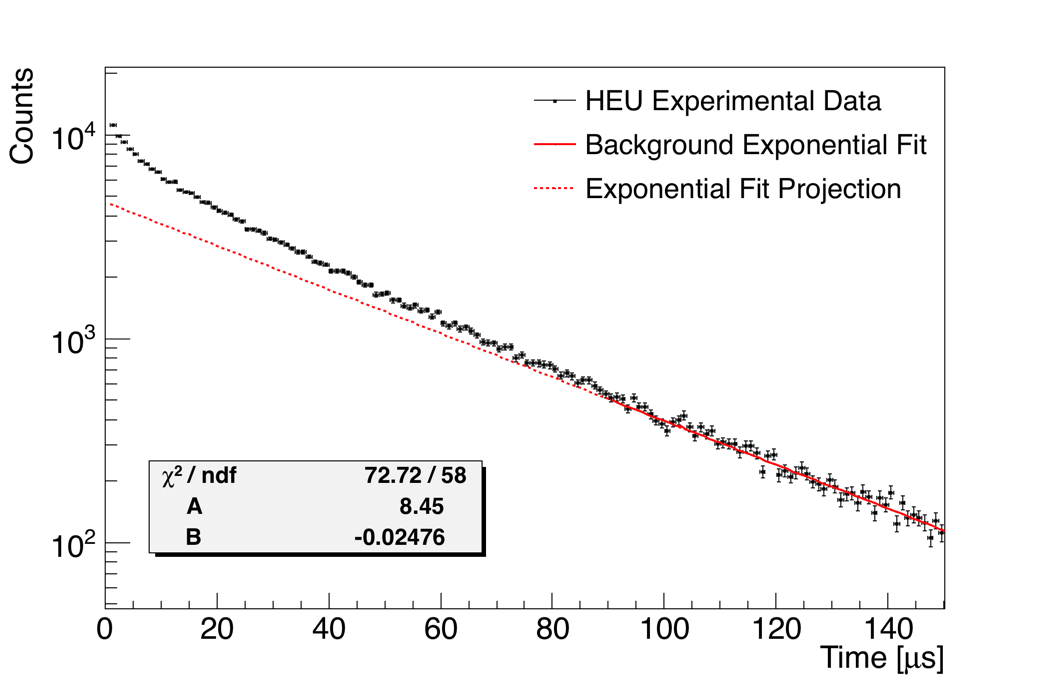} \\
    \mysubfigimg[width=\linewidth]{\hspace{0.5cm}(b)}{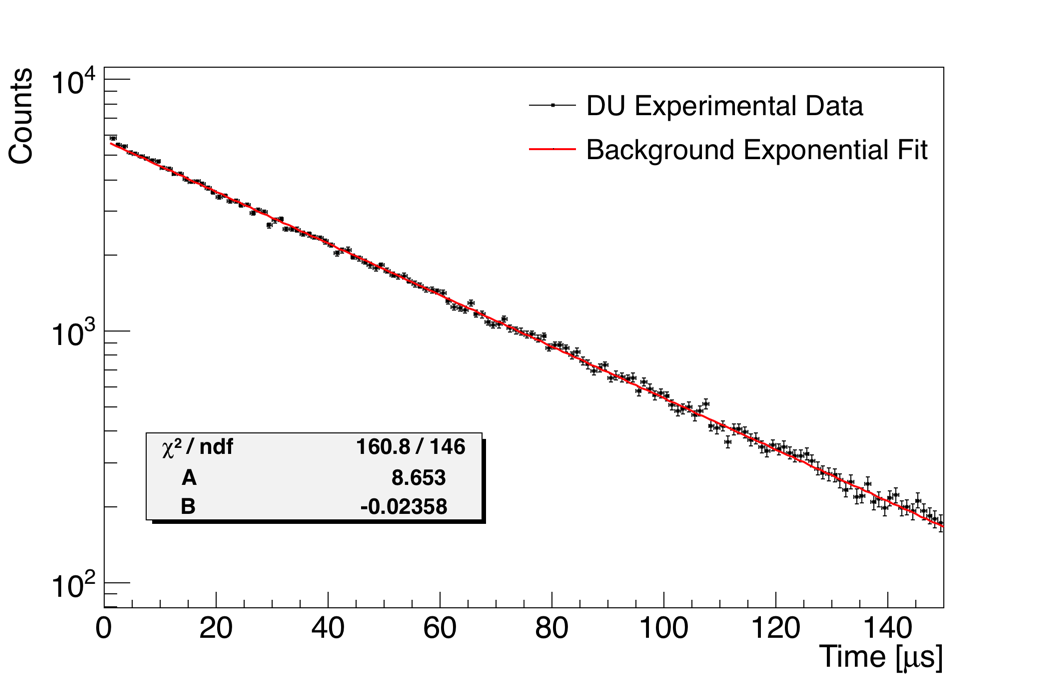}  \\
  \end{tabular}
  \caption{Experimental recoil-capture coincidence time distributions in the composite detector for (a) HEU and (b) DU. Each curve is fitted with an exponential function representing the expected contribution of background accidentals. For HEU, the $\chi^2$ value applies only to the range of times beyond 90~\textmu s. For DU, the $\chi^2$ value applies to the full range.}
  \label{fig:thermTime}
\end{figure}

Examination of the time distribution of coincident recoil-capture pulses can provide additional insight into the observed light output spectra. While the simulated light output response for delayed neutrons from $^{238}$U is largely restricted to below 400~keVee, the experimentally measured distribution extends to higher light outputs. Since the delayed neutrons do not have enough energy to produce higher light output pulses in the detector (and a large fraction cannot even produce a pulse above a $\sim$100~keVee detection threshold), this suggests either that there is a higher rate of delayed fission than predicted by tabulated nuclear data, or that the capture-gated light output distribution is dominated by gamma-ray accidentals from background or passive emission by the DU object. If accidentals dominate, then the distribution $I(t)$ of time differences between two adjacent pulses is governed by the general expression
\begin{equation}
    I(t) = r \exp(-rt),
    \label{eq:TimeDist}
\end{equation}
where $t$ is the time between two pulses, and $r$ is the rate of accidentals. In the scenario where there are very few true thermalization events, Eq.~(\ref{eq:TimeDist}) predicts that the time distribution should exhibit a simple exponential decay shape. Fig.~\ref{fig:thermTime} shows the experimental recoil-capture coincidence time distributions for HEU and DU, which have been fitted with an exponential function representing the expected contribution of background accidentals. 

In both cases, the exponential function was fitted to the time window beyond 90~\textmu s, where the contribution of true neutron thermalization events is negligible (less than 1\% of the total distribution). The model is then extended back over the range of possible neutron thermalization times. In the case of DU, the exponential function is consistent with the entire distribution, suggesting that few, if any, of the pulses preceding neutron capture events are caused by thermalization of delayed neutrons. In contrast, the time distribution for HEU departs significantly from a simple exponential shape in the neutron thermalization window, and the point where this deviation becomes noticeable is near the maximum thermalization time of 76~\textmu s predicted by simulation. After subtracting the background exponential fit, 98.5\% of capture-gated recoil pulses occur within the 76~\textmu s time window, in close agreement with the simulated result of 99\%. Furthermore, the decay constant of the exponential fit is very similar for each data set, which suggests a common cause of accidentals, such as gamma-ray background. The idea that gamma-ray accidentals dominate the recoil-capture coincidences for DU above 400~keVee was validated by examining the overall recoil rates in the composite detector. Details on this analysis are presented in the Supplementary Information. The significant difference in the time-to-capture curves for HEU and DU suggest that this signal may provide yet another means for performing isotopic discrimination.

\begin{figure}[!ht]
  \centering
  \begin{tabular}{@{}p{1.0\linewidth}@{}p{1.0\linewidth}@{}}
    \mysubfigimg[width=\linewidth]{\hspace{5cm}\color{black}(a) HEU}{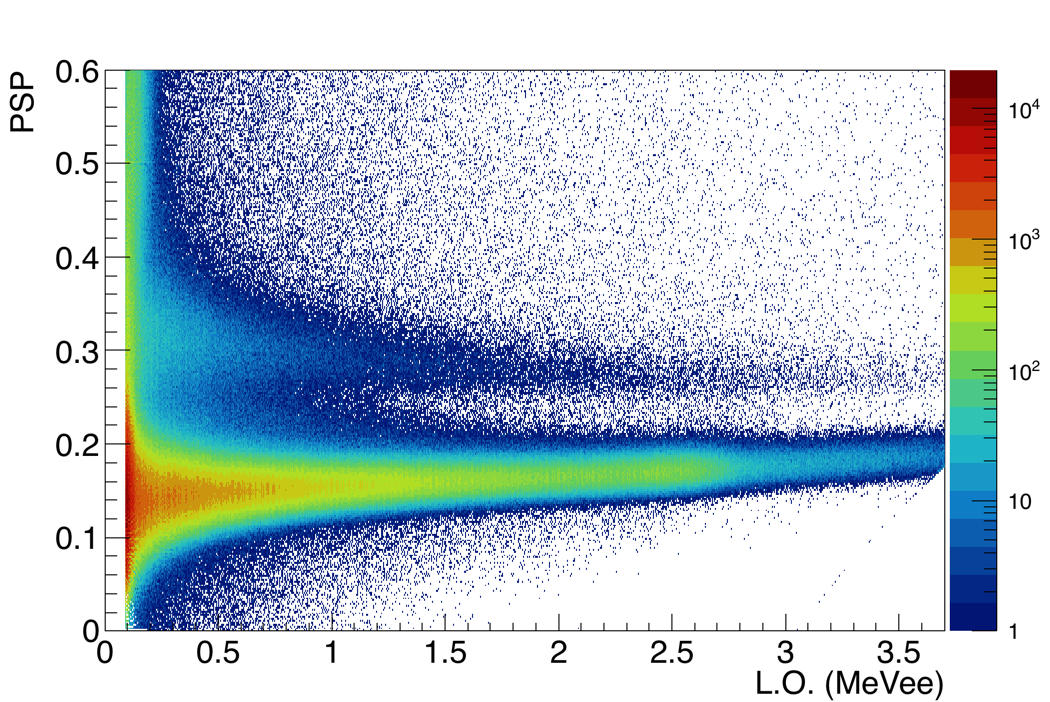} \\
    \mysubfigimg[width=\linewidth]{\hspace{5cm}\color{black}(b) DU}{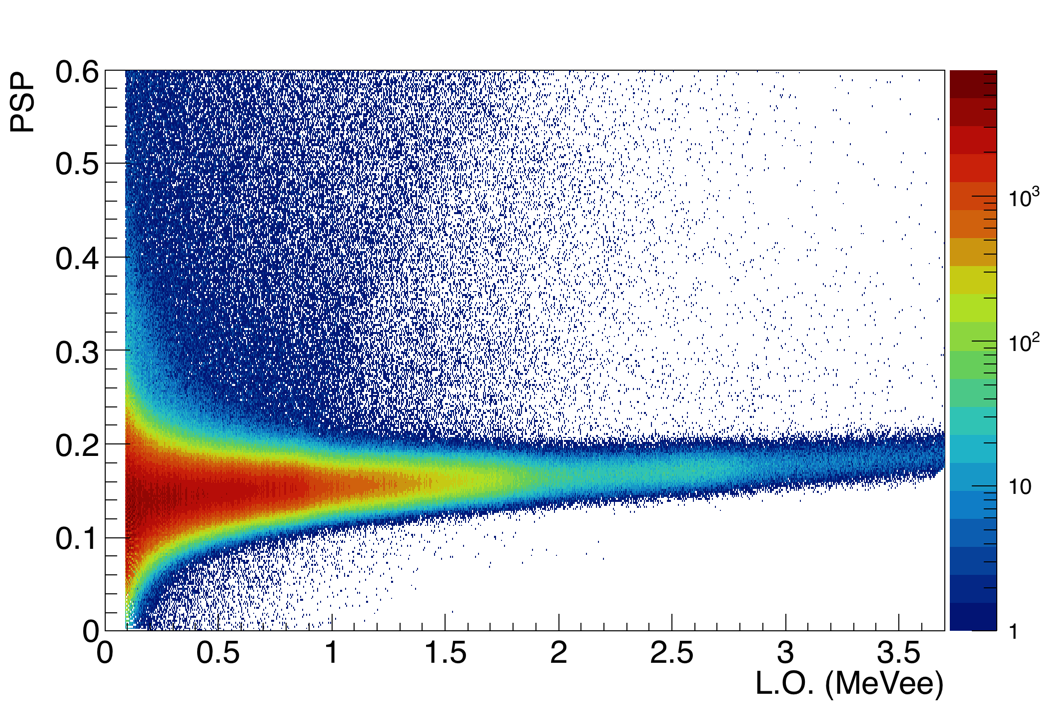}  \\
  \end{tabular}
  \caption{$PSP$ and light output distributions measured in EJ309 for (a)~HEU and (b)~DU.}
  \label{fig:PSP_EJ}
\end{figure}

Recoil-based organic liquid scintillators are poorly suited to detecting lower-energy primary delayed neutrons because those neutrons are unlikely to produce a response above a detection threshold of $\sim$100~keVee. However, higher-energy prompt neutrons from delayed fission can be easily detected, as they are far more likely to produce a pulse above the threshold. Fig.~\ref{fig:PSP_EJ} shows the $PSP$ and light output distributions measured by the EJ309 detector for HEU and DU, respectively. In the HEU data, a fast neutron recoil region around $PSP$=0.28 is readily apparent. However, this feature is entirely absent from the DU distribution. This stark contrast provides convincing evidence that the presence or absence of high-energy fission neutrons in the delayed neutrons spectrum can be used to perform isotopic discrimination.

Coincidence measurements were also used to detect the presence of delayed neutron-induced fission events. Coincidence time distributions were recorded for HEU and DU using two different detector pairings: composite-EJ309 and composite-NaI(Tl). In each case, the coincidence event rate recorded during the neutron generator off cycles was compared with the rate observed during passive measurement for each uranium object. A summary of the measured coincidence rates for each detector pairing and scenario is presented in Table~\ref{table:CoincidenceAll}. Example coincidence distributions are included in the Supplementary Information.

\begin{table}[h]
\parbox{1.0\linewidth}{
\caption{Measured coincidence rates for HEU and DU for composite-EJ309 and composite-NaI(Tl) detector pairings. For each material, the coincidence rate observed during delayed signal measurements was compared to the coincidence rate for passive measurements.}
\centering
\begin{tabular}{ccc}
\hline
\hline
 & HEU & DU \\
\hline
\textbf{Composite-EJ309} & & \\
Passive Rate (s$^{-1}$) & \ \ 55.87 $\pm$ 0.70 \ \  & 50.30 $\pm$ 0.34  \\
Active Rate (s$^{-1}$) & \ \ 89.17 $\pm$ 0.25 \ \  & 58.90 $\pm$ 0.16  \\
\% Change & \ \ 60\% \ \  & 17\%  \\
\textbf{Composite-NaI(Tl)} & & \\
Passive Rate (s$^{-1}$) & \ \ 1.70 $\pm$ 0.25 \ \  & 2.62 $\pm$ 0.14  \\
Active Rate (s$^{-1}$) & \ \ 3.78 $\pm$ 0.10 \ \  & 2.91 $\pm$ 0.07  \\
\% Change & \ \ 122\% \ \  & 11\%  \\
\hline
\hline
\end{tabular}
\label{table:CoincidenceAll}
}
\end{table}

Because both samples contain $^{238}$U, which undergoes spontaneous fission, the comparison between HEU and DU is not as simple as noting the presence or absence of coincident radiation from fission. While the relative change in the total coincidence rate after interrogation (when delayed neutrons are present) is much greater for HEU, suggesting that much of the change is due to delayed neutron-induced fission events, this interpretation must be weighted against a number of complicating factors. 

Based on further simulation and analysis, which is presented in detail in the Supplementary Information, the measured passive coincidence rate for each material is too high to be attributed to spontaneous fission alone. The increased passive coincidence rates may be partly attributed to a high gamma-ray background at the DAF, as well as activation caused by the DT generator. Furthermore, the HEU object can be fissioned by thermal neutrons, which can cause additional coincidence events. Combined with natural background, the presence of neutron-emitting calibration sources, such as AmBe, in the room during the experiment means that thermal neutron induced fission may also have been a factor.

With the goal of eliminating events that were not caused by fission, the coincidence rates were reexamined while only accepting neutron recoil pulses from the EJ309 detector. Examples of the resulting coincidence time distributions are included in the Supplementary Information. Table~\ref{table:CoincidenceNGamma} presents a summary of the experimentally measured coincidence rates when the neutron recoil criterion is applied. When coincidence events are required to contain at least one neutron interaction, the differences between the HEU and DU coincidence rates become much more pronounced. 

\begin{table}[h]
\parbox{1.0\linewidth}{
\caption{Experimental fission tagging rates for HEU and DU using the composite-EJ309 detector pairing, with only neutron events accepted for the EJ309 detector.}
\centering
\begin{tabular}{ccc}
\hline
\hline
 & HEU & DU \\
\hline
Passive Rate (s$^{-1}$) & \ \ 0.442 $\pm$ 0.054 \ \  & 0.035 $\pm$ 0.007  \\
Active Rate (s$^{-1}$) & \ \ 9.14 $\pm$ 0.07 \ \  & 0.079 $\pm$ 0.005  \\
\% Change & \ \ 1968\% \ \  & 126\%  \\
\hline
\hline
\end{tabular}
\label{table:CoincidenceNGamma}
}
\end{table}

In the case of DU, the coincidence rate more than doubles, even though delayed neutrons are not expected to cause an increase in coincidences due to fission. However, this is most likely due to an increased number of pileup events in the neutron recoil region for EJ309, as Fig.~\ref{fig:PSP_EJ}(b) suggests. While the passive coincidence rate for HEU is still quite high relative to expectation, it is much lower than when all events are considered. Given the presence of other neutron sources in the experimental space, it is also reasonable to assume that some of the discrepancy is accounted for by additional fissions caused by thermal neutrons. Most notably, the overall coincidence rate for HEU after interrogation is more than 100 times higher than the rate for DU, and the change in the HEU coincidence rate between the passive and active measurements is very significant, increasing by more than a factor of 20. This is consistent with the expectation that delayed neutrons will induce additional fission at a much greater rate in HEU than DU, providing the basis for discrimination.

\begin{figure}[!ht]
\centering
\includegraphics[width=\linewidth]{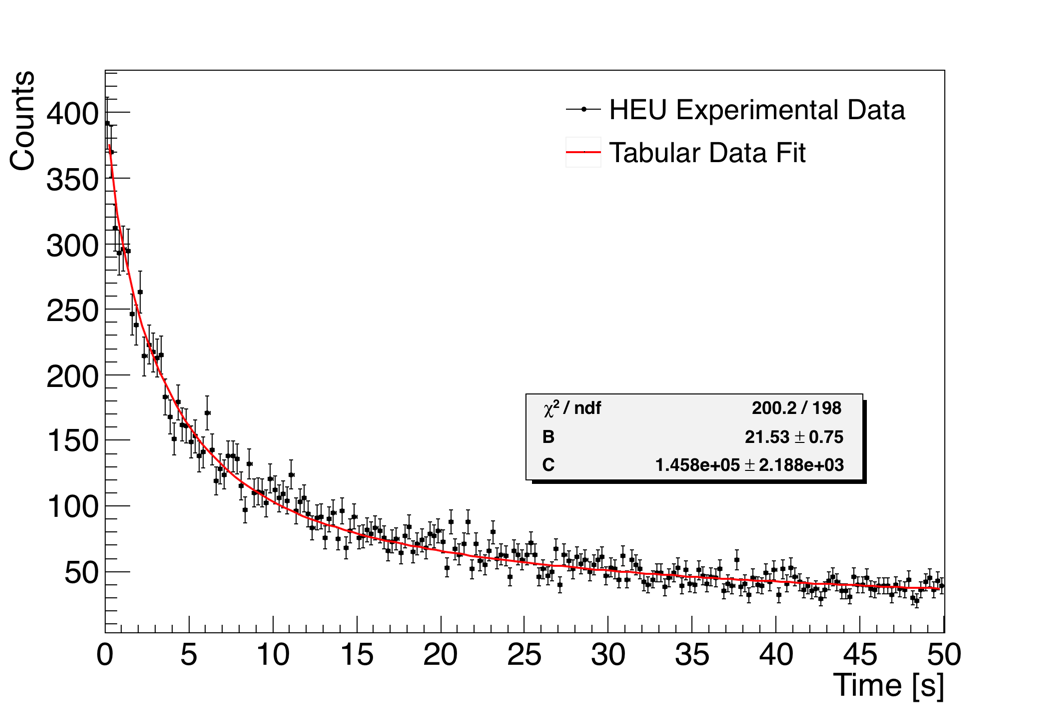}\hspace{1cm}
\caption{Time distribution of delayed coincidence events in HEU, measured using the composite-EJ309 detector pairing. Only neutron recoil pulses were accepted from the EJ309 detector. The fit is based on tabular nuclear data and parameterized with only a scaling factor (C) and constant background term (B).}
\label{fig:heu_coincidenceRate_ngamma}
\end{figure}

Furthermore, the time evolution of the rate of coincidence events from fission can provide valuable information on the $^{235}$U content of the test material. Because the delayed neutron groups for each uranium isotope constitute a unique set of decay time constants, the overall delayed neutron time emission profile can be used to discriminate between isotopes and infer enrichment. Delayed neutron-induced fission events occur on the same timescale as their delayed neutron precursors, so coincidence-based measurements of the rate of delayed fission events should exhibit the same temporal shape predicted for delayed neutron emission. Fig.~\ref{fig:heu_coincidenceRate_ngamma} shows the time distribution of coincidence events for HEU in the period after the neutron generator has been turned off.  

The composite-EJ309 detector pairing was used again, but while accepting only neutron recoil events from the EJ309 detector. The experimental data are fitted with a parameterized model based on a six-group superposition of delayed neutrons, whose decay constants are obtained from tabular nuclear data; the procedure is described in depth in Ref.~\cite{Nattress2018}. The experimental results show close agreement with the model ($\chi^2 = 200.2/198$), which confirms that the coincidence events are caused by delayed neutron-induced fission and suggests that discrimination based on the delayed neutron time emission profile can also be performed using fast neutron measurements, provided that the material is fissionable by lower-energy delayed neutrons. The coincidence-based approach presented here could supplement the methods described in Ref.~\cite{Nattress2018}, providing an additional point of distinction between isotopes.

In conclusion, we have demonstrated two measurement methods for differentiating the components of delayed neutron signals in bulk samples of SNM based on their origin. Through fission radiation coincidence counting and spectroscopic neutron energy measurements, we have shown a significant refinement in the ability to capture information on delayed neutron-induced fission as a means for performing isotopic identification. For fission rate measurements, we found that high background gamma-ray rates hindered the usefulness of NaI(Tl) detectors, and detectors with the ability to discriminate fast neutron recoils were necessary to isolate the fission signature. For fissionable materials with large differences in fission cross-section at typical delayed neutron energies, such as $^{235}$U and $^{238}$U, we show that these types of measurements are sufficient to successfully perform isotopic discrimination. Further refinement of the capture-gated neutron spectroscopy technique to detect small changes in the delayed neutron energy signature would provide even greater precision in differentiating materials. The measurement approaches presented here have the potential to complement existing delayed neutron analysis techniques, and when employed in concert with methods that focus on time-dependent signatures, they may lead to even greater accuracy in SNM characterization.


\section*{Acknowledgments}
The authors would like to thank J. Mattingly of North Carolina State University and J. Hutchinson of Los Alamos National Laboratory for their assistance in organizing and executing the experimental campaign at the DAF. This work was supported by the U.S. Department of Homeland Security under Grant Award No. 2014-DN-077-ARI078-02 and 2015-DN-077-ARI096 and by the Consortium for Verification Technology and Consortium for Monitoring, Verification and Technology under U.S. Department of Energy National Nuclear Security Administration award numbers DE-NA0002534 and DE-NA0003920, respectively. The research of J. Nattress was performed under appointment to the Nuclear Nonproliferation International Safeguards Fellowship Program sponsored by the National Nuclear Security Administration's Office of International Safeguards (NA-241).

\bibliography{DAFBib2018,rev-tex-custom}

\end{document}


\beginsupplement
\maketitle

\vspace{-20pt}
\subsubsection*{\centering{Supplementary Note 1: Details on Uranium Test Objects}}

A set of concentric HEU hemispheres known as the Rocky Flats shells was used as the HEU object. The Rocky Flats shells have a bulk density of 18.664~g/cm$^3$ and an isotopic content of 93.16\% $^{235}$U, 5.35\% $^{238}$U, and less than 2\% of other isotopes of uranium~\cite{Rothe1997}. Shells 01--24 were assembled into a spherical object with an approximate mass of 13.8~kg. Detailed information on the masses and dimensions of the individual shells can be found in Ref.~\cite{Rothe1997}. The depleted uranium object also consisted of a set of hemispherical shells arranged to form a sphere. The mass of the DU object was 12.8~kg, which was the closest approximation to the HEU object mass that could be achieved with the available shell configurations.

\subsubsection*{\centering{Supplementary Note 2: Composite Detector Design and Principle of Operation}}

\begin{figure}[!ht]
  \centering
  \begin{tabular}{@{}p{0.42\linewidth}@{}p{0.48\linewidth}@{}}
    \simsubfigimg[width=0.8\linewidth]{(a)}{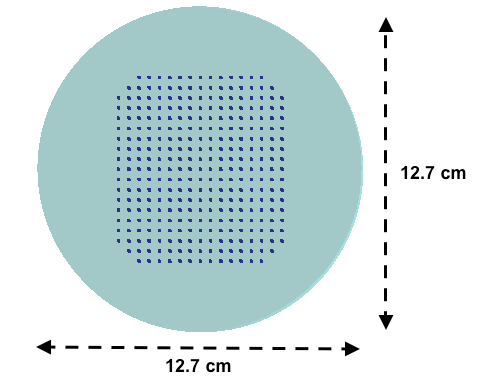} &
    \simsubfigimg[width=0.8\linewidth]{(b)}{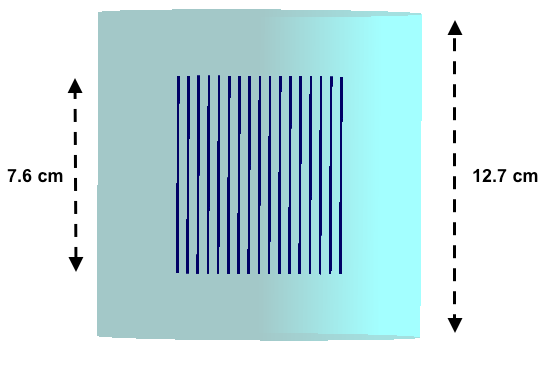} \\
  \end{tabular}
  \caption{Geant4 rendering of the composite detector geometry: (a)~top and (b)~side view} 
\label{fig:simDet}
\end{figure}

The composite detector is a larger version of the prototype described in Ref.~\cite{Mayer2015}, and consists of an array of 1$\times$1$\times$76~mm$^3$ GS20 lithium glass square rods embedded in a cylindrical matrix of scintillating polyvinyl toluene (PVT) with a height and diameter of 12.7~cm. The GS20 glass is 6.6\% lithium by weight, and is enriched to approximately 95\% $^6$Li. A diagram of the composite detector design is shown in Fig.~\ref{fig:simDet}.

The principal detection mechanism for the composite detector is neutron capture by $^6$Li, which has a reaction $Q$-value of 4.8~MeV and releases a triton and an alpha particle. The short range of the heavy charged particle products means that they generally deposit most of their energy in the lithium-doped glass, which possesses very different scintillation properties from the PVT plastic. As a result, neutron capture events are easily distinguishable by both a characteristic pulse shape and the characteristic $Q$-value of the reaction. The PVT matrix surrounding the lithium glass rods serves a dual purpose. Not only does it increase the capture efficiency of the detector by moderating the incident neutrons, but the scintillation response of the PVT to proton recoils in the neutron thermalization process provides a signal whose magnitude is correlated to the incident neutron energy~\cite{Shi2016}. The recoil pulses from neutron thermalization are also correlated in time with the subsequent capture pulses. Though there can be significant variation in the time separation between pulses due to thermal neutron diffusion in the PVT, the average time to capture ranges from a few \textmu s to about 10~\textmu s, depending on the detector design and size. By exploiting this time coincidence between a capture pulse and the preceding proton recoil pulse, spectroscopic neutron energy analysis can be performed~\cite{Nattress2016}.

\subsubsection*{\centering{Supplementary Note 3: Details on Data Acquisition}}

The detector electronics included Hamamatsu model R6527 and R6231 photomultiplier tubes (PMT), which were coupled to the composite detector (biased at 1750~V) and one of the EJ309 detectors (biased at 900~V), respectively. The other EJ309 detector was coupled to a 7.6-cm diameter Bicron PMT (biased at 1200~V), and the NaI(Tl) was a Gamma Spectacular GS-2020 integrated detector and PMT unit (biased at 750~V). Each detector was powered using a CAEN DT5533 high-voltage power supply, and the output was digitized using a CAEN DT5730 14-bit, 500-MHz desktop waveform digitizer with digital pulse processing-pulse shape discrimination (DPP-PSD) firmware. Data acquisition and storage was performed using CAEN Multi-Parameter Spectroscopy Software (CoMPASS)~\cite{COMPASS}. For each waveform, short-gate ($Q_{\mathrm{short}}$) and long-gate ($Q_{\mathrm{long}}$) charge integrals were recorded to provide the basis for pulse shape discrimination. The integration boundary parameters were defined relative to $t_{\mathrm{start}}$, the start of the waveform determined by the leading edge trigger time in the digitizer. $Q_{\mathrm{short}}$ was integrated from $t_{\mathrm{start}}-t_{\mathrm{offset}}$ to $t_{\mathrm{start}}+t_{\mathrm{short}}$, and $Q_{\mathrm{long}}$ was integrated from $t_{\mathrm{start}}-t_{\mathrm{offset}}$ to $t_{\mathrm{start}}+t_{\mathrm{long}}$, where $t_{\mathrm{offset}}$ is the offset time prior to the start of the waveform, and $t_{\mathrm{short}}$ and $t_{\mathrm{long}}$ are the endpoints for the short-gate and long-gate integration windows, respectively. The integration parameters were optimized for each detector prior to the experiment, and are presented in Table~\ref{table:PSDparams}.

\begin{table}[ht]
\parbox{1\linewidth}{
\caption{Waveform integration parameters.}
\centering
\begin{tabular}{cccc}
\hline
\hline
Detector & $t_{\mathrm{offset}}$~(ns) & $t_{\mathrm{short}}$~(ns) &  $t_{\mathrm{long}}$~(ns) \\
\hline
Composite & \ \ 24 \ \  & 36 & \ \ 376  \ \  \\
EJ309 - 1 & \ \ 24 \ \  & 36 & \ \ 376 \ \   \\
EJ309 - 2 & \ \ 50 \ \  & 50 & \ \ 350 \ \   \\
NaI(Tl) & \ \ 24 \ \  & 176 & \ \ 776 \ \   \\
\hline
\hline
\end{tabular}
\label{table:PSDparams}
}
\end{table}

\subsubsection*{\centering{Supplementary Note 4: Delayed Neutron Spectrum Simulation Methods}}

In the simulation conducted using the MCNP-Polimi code~\cite{Pozzi2012}, spherical HEU and DU objects of the approximate size and mass of the experimentally measured assemblies were interrogated with 14.1-MeV neutrons, and the time of generation and initial energy of emitted neutrons were recorded upon their arrival in a liquid scintillation detector. The prompt and delayed neutron spectra were separated using a simple time threshold, where neutrons arriving at the detector within 10~\textmu s of the generation of the source particle were vetoed. The time threshold was chosen based on the simulated time distribution of prompt neutrons. A cutoff of 10~\textmu s excludes 99.99\% of the prompt neutrons, while the vast majority of delayed neutrons arrive much later than 10~\textmu s after their parent fission event. Within the delayed neutron data, primary delayed neutrons and prompt neutrons from delayed fission were differentiated by tracking their individual histories in the MCNPX-PoliMi collision file output. Based on the particle and generation numbers of each delayed neutron, the corresponding parent fission event was located within the fission chain in the uranium object. If the time delay between the parent fission and detection of the delayed neutron was greater than 10~\textmu s, the event was categorized as a primary delayed neutron. Otherwise, it was considered to be a prompt neutron from delayed fission. This analysis method allowed for estimation of the relative contribution of each type of delayed neutron to the overall delayed spectrum emitted by each object. 

\subsubsection*{\centering{Supplementary Note 5: A Method for Isotope Discrimination Based on Capture-Gated Spectrum Shape}}

\begin{figure}[h]
\centering
\includegraphics[width=0.5\linewidth]{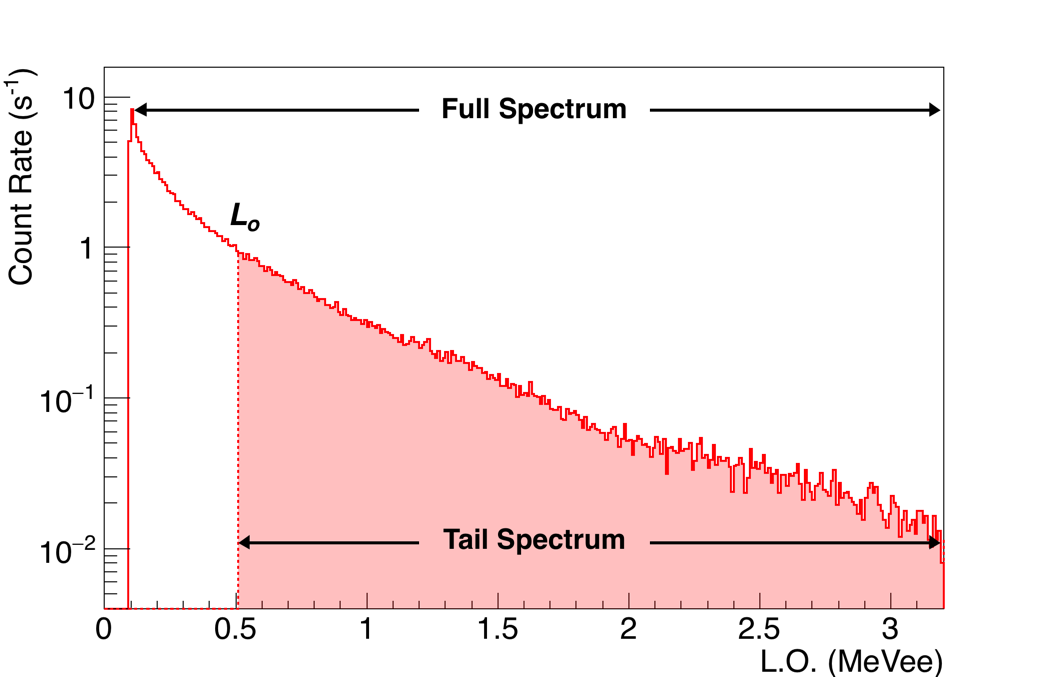}
\caption{Diagram of the proposed integration regions for determining the characteristic spectral shape ratio. The example spectrum shown is for HEU.}
\label{fig:HEU_shape}
\end{figure}

The capture-gated event rates above 400 keVee are 58.95~s$^{-1}$ for HEU and 18.76~s$^{-1}$ for DU. HEU and DU could be distinguished based on rate comparison, but such a method relies on the knowledge of sample mass. A more robust isotope discrimination method would be based on differences in the shapes of the neutron energy spectra. One possibility is to use a simple ratio of two different regions of the energy spectrum, which would provide a characteristic number associated with the proportion of prompt neutrons from delayed fission in the spectrum (and thus an indicator of enrichment).  Fig.~\ref{fig:HEU_shape} shows a diagram of our proposed metric, which is the ratio of the integral of the high-energy tail of the neutron spectrum to the integral of the entire spectrum. We refer to this ratio as $R$, and to its standard deviation as $\sigma$. A figure of merit (FOM) for discrimination could be determined using the expression
\begin{equation}
    \mathrm{FOM} = \frac{\overline{R}_{\mathrm{HEU}} - \overline{R}_{\mathrm{DU}}}{\sigma_{R_{\mathrm{HEU}}} + \sigma_{R_{\mathrm{DU}}}} ,
\end{equation}
similarly to a standard method used for evaluating pulse-shape-discrimination performance of scintillators.

The ratio $R$ is determined by the choice of parameter $L_{\mathrm{o}}$, which is the light-output threshold separating the high-energy tail from the rest of the spectrum. The threshold $L_{\mathrm{o}}=515$~keVee shown in Fig.~\ref{fig:HEU_shape} is the result of optimizing for the maximum separation between the ratios for HEU and DU along with the minimum statistical error. The maximum ratio separation occurs at 515~keVee, where the HEU ratio, $R_{\mathrm{HEU}}$, is 0.302 and the DU ratio, $R_{\mathrm{DU}}$, is 0.178. While the minimum statistical error of 1.1\% occurs a higher $L_{\mathrm{o}}$ value of about 700 keVee, the error at 515~keVee is only very slightly higher at 1.24\%. As such, the marginal loss in ratio separation by moving the $L_{\mathrm{o}}$ value closer to 700~keVee would be greater than the marginal improvement in statistical error. 

The shape-based discrimination method appears promising when applied to the experimental data. Using the $L_{\mathrm{o}}$ value of 515~keVee, the tail-spectrum count rates were 45.8~s$^{-1}$ and 12.2~s$^{-1}$ for HEU and DU respectively, while the total-spectrum count rates were 151.7~s$^{-1}$ and 68.8~s$^{-1}$. These experimental count rates allow for 3-$\sigma$ discrimination of the HEU and DU characteristic ratios within 2 seconds of measurement time after the neutron generator has been turned off. Total measurement time would be just over one minute, as the one minute interrogation time is necessary to build up the delayed neutron populations.

\subsubsection*{\centering{Supplementary Note 6: Total Recoil Rate Analysis in the Composite Detector}}

The overall rate of recoil pulses recorded by the composite detector between generator runs for DU is consistent with the assertion that the events in the DU capture-gated spectrum above 400~keVee are caused by gamma-ray accidentals in the thermalization window. When the generator is turned off following interrogation, the observed rate of recoil pulses is $\sim$2.1$\times 10^4$ s$^{-1}$. At this rate, the average time between events is 48~\textmu s, and the probability of observing a random recoil pulse within the 76~\textmu s time window preceding a capture pulse is 79.5\%. In the experimental results, 74.0\% of recorded capture events are accompanied by a preceding pulse within 76~\textmu s. 

The observed rate of events within the thermalization window is close to the prediction if all thermalization candidate pulses were truly from gamma-ray accidentals, but it is still somewhat too low. This may be explained by the effects of delayed neutron contributions to the recoil $PSP$ region and pileup pulses in the neutron capture region. Specifically, if the recoil rate contains significant contributions from delayed neutrons, the rate of events in the thermalization window will be higher than expected for random gamma-ray accidentals because neutron recoils are much more likely to be followed by a subsequent capture event. Furthermore, misclassified pileup pulses in the capture $PSP$ region are less likely to be preceded by a neutron recoil event. This results in a situation where the random gamma-ray rate, and thus the probability of an event within the thermalization time window, is slightly overestimated.

To correct for the effects of delayed neutron recoils on the estimation of the gamma-ray accidental rate, a 400~keVee light output threshold was applied, effectively removing neutron contributions from the spectrum. The threshold was determined based on the simulated light output response of the composite detector to delayed neutrons, as shown in Fig.~3. The overall rate of recoil events above the threshold was 5.6$\times 10^3$ s$^{-1}$, resulting in an average time between events of 177~\textmu s and a 34.9\% probability of observing a random event within the 76~\textmu s thermalization window. In the experimental data, 33.9\% of capture events were accompanied by a preceding recoil event within 76~\textmu s, which strongly suggests that gamma-ray accidentals contribute significantly to the the rate of recoil-capture coincidence events in the DU light output spectrum (Fig.~5) above 400~keVee. Table~\ref{table:AccidentalRates} provides a summary of the recoil rates, probability of gamma-ray accidentals in the thermalization window, and comparison to experimental data, both with and without application of the 400~keVee light output threshold.

\begin{table}[h]
\parbox{1.0\linewidth}{
\caption{Summary of overall recoil rates in the composite detector and probability of gamma-ray accidentals within the 76 \textmu s thermalization window. Predicted probabilities based on the overall rate are compared to experimental results.}
\centering
\begin{tabular}{ccc}
\hline
\hline
 & \multirow{ 2}{*}{All Events} & Events \\
 & & $>$400~keVee \\
\hline
& & \\
Recoil Rate (s$^{-1}$) & 2.1$\times 10^4$ & 5.6$\times 10^3$ \\
& & \\
Average Time & \ \ \multirow{ 2}{*}{48} \ \  & \multirow{ 2}{*}{177}  \\
Between Events (\textmu s) & \ \  \ \  &   \\
& & \\
Probability of Event & \ \ \multirow{ 2}{*}{79.5} \ \  & \multirow{ 2}{*}{34.9}  \\
Within 76 \textmu s (\%)& \ \  \ \  &   \\
& & \\
\% of Captures with Preceding & \ \ \multirow{ 2}{*}{74.0} \ \  & \multirow{ 2}{*}{33.9}  \\
Event Within 76 \textmu s & \ \  \ \  &   \\
\hline
\hline
\end{tabular}
\label{table:AccidentalRates}
}
\end{table}

\subsubsection*{\centering{Supplementary Note 7: Coincidence Time Distributions}}

\begin{figure}[!ht]
  \centering
  \begin{tabular}{@{}p{0.48\linewidth}@{}p{0.48\linewidth}@{}}
    \simsubfigimg[width=\linewidth]{\hspace{1.2cm}(a)}{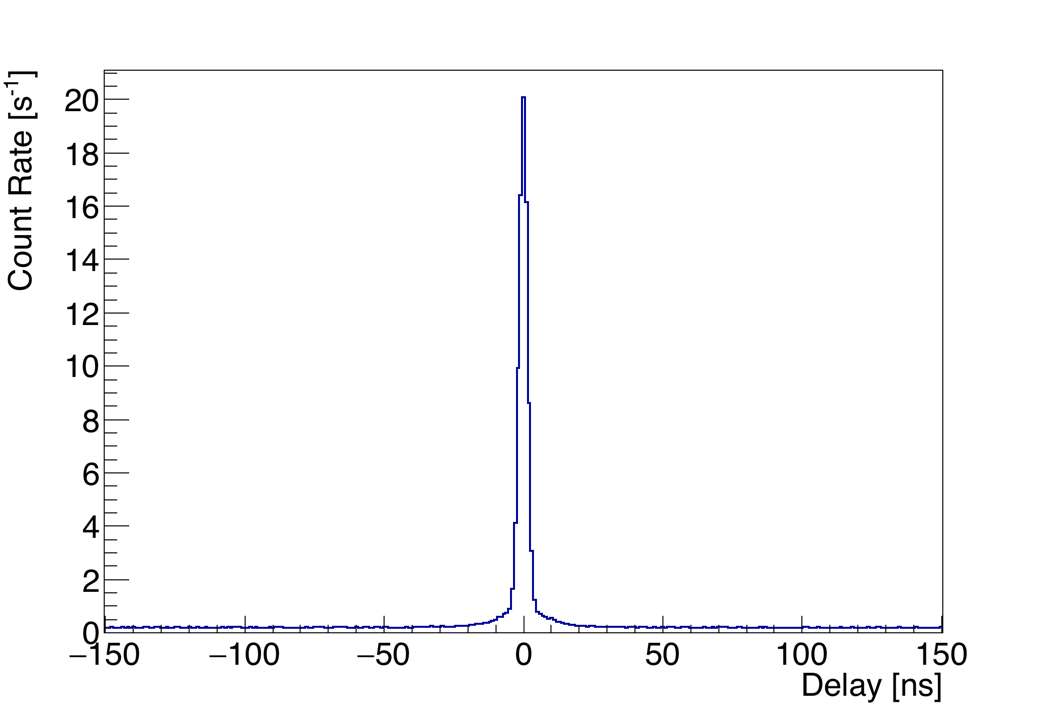} &
    \simsubfigimg[width=\linewidth]{\hspace{1.2cm}(b)}{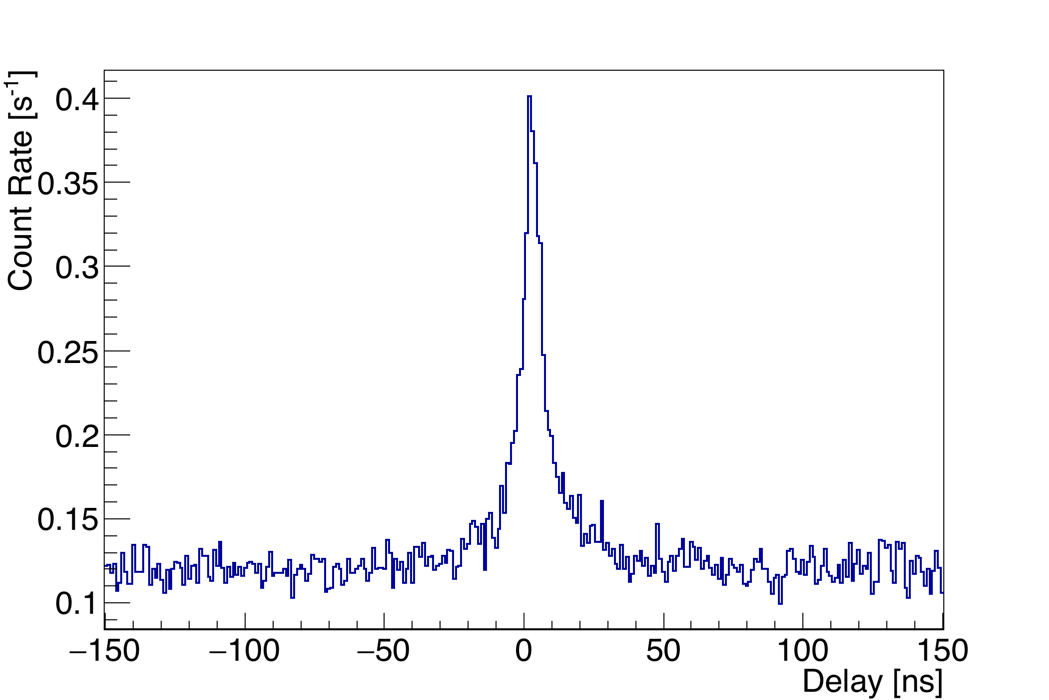} \\
  \end{tabular}
  \caption{Coincidence time distributions for HEU measured during the delayed neutron time window based on (a)~composite-EJ309 and (b)~composite-NaI(Tl) detector pairings.}
  \label{fig:Coincidence_HEU}
\end{figure}

Example coincidence time distributions measured during the delayed neutron window are shown in Fig.~\ref{fig:Coincidence_HEU} for the composite-EJ309 and composite-NaI(Tl) detector pairings, respectively. The prominent coincidence peaks provide a clear indication that fission events continue to take place during the delayed neutron window.

\subsubsection*{\centering{Supplementary Note 8: Analysis of Passive Coincidence Rates}}

The measured passive coincidence rates for each material are too high to be attributed to spontaneous fission alone. $^{238}$U has a specific activity of 12.44 MBq/kg, and a spontaneous fission probability of $5.4 \times 10^{-7}$ per decay. Thus, based on the material composition and total mass, the expected spontaneous fission rate for the DU and HEU objects would be 87 fission/s and 5 fissions/s, respectively. MCNPX simulations of the experimental configuration show that the probability of a fission event producing observable coincident pulses in the composite and EJ309 detectors is 0.3\% for HEU and 0.053\% for DU. For coincidence pulses in the composite and NaI(Tl) detectors, the probabilities are 0.08\% and 0.0006\% for HEU and DU, respectively. Table~\ref{table:CoincidenceSimAll} shows the expected spontaneous fission (SF) tagging rates for each object based on the calculated spontaneous fission rates and simulated tagging probabilities. 

\begin{table}[h]
\parbox{1.0\linewidth}{
\caption{Simulated spontaneous fission (SF) rates, tagging probability, and expected tagged event rates for HEU and DU using each detector pairing.}
\centering
\begin{tabular}{ccc}
\hline
\hline
 & HEU & DU \\
\hline
\textbf{Composite-EJ309} & & \\
SF Rate (s$^{-1}$) & \ \ 5 \ \  & 87  \\
Tagging Probability  & \ \ 0.30\% \ \  & 0.053\%  \\
Tagged Event Rate (s$^{-1}$) & \ \ 0.015 \ \  & 0.046  \\
\textbf{Composite-NaI(Tl)} & & \\
SF Rate (s$^{-1}$) & \ \ 5 \ \  & 87  \\
Tagging Probability  & \ \ 0.08\% \ \  & 0.006\%  \\
Tagged Event Rate (s$^{-1}$) & \ \ 0.004 \ \  & 0.0052  \\
\hline
\hline
\end{tabular}
\label{table:CoincidenceSimAll}
}
\end{table}

In both the case of HEU and DU, the expected tagging rates are several orders of magnitude lower than the experimentally observed coincidence rates. Further investigation of the coincidence pulses showed that they were consistent with true recoil events and not caused by spurious sources such as electronic noise. The measurement environment at Device Assembly Facility is complex, with many other objects near the experimental setup being subjected to bombardment by 14.1-MeV neutrons, so the high coincidence rates may be partially explained by gamma-ray radiation from neutron activation products. In addition to any fission byproducts within the test objects, the entire experimental array was placed on a carbon steel table, so coincident gamma rays from $^{60}$Co (produced by neutron irradiation of Fe) may be a contributing factor. While MCNPX simulations show the likelihood for crosstalk to be low ($<$2\% for the composite-EJ309 pairing, negligible for composite-NaI(Tl)), a high background gamma-ray flux due to activation following the neutron generator operation could lead to an appreciable coincidence signal in the composite-EJ309 detector pairing due to crosstalk. 

With respect to thermal neutron background, MCNPX simulations show that thermal neutrons incident on the HEU object are about 70\% as likely to produce coincident pulses as spontaneous fission events within the object when averaged over all directions, and much more likely to produce coincident pulses when incident near the EJ309 and composite detectors. As such, even a small thermal neutron flux could cause an appreciable change to the measured coincidence rate for HEU. Given that neutron-emitting calibration sources were present in the room during the experiment, it is possible that thermal neutron induced fission played a part in elevating the measured passive coincidence rates. 

\subsubsection*{\centering{Supplementary Note 9: Coincidence Distributions with Neutron Pulse Shape Discrimination Cut}}

\begin{figure}[!ht]
  \centering
  \begin{tabular}{@{}p{0.48\linewidth}@{}p{0.48\linewidth}@{}}
    \simsubfigimg[width=\linewidth]{\hspace{1.2cm}(a)}{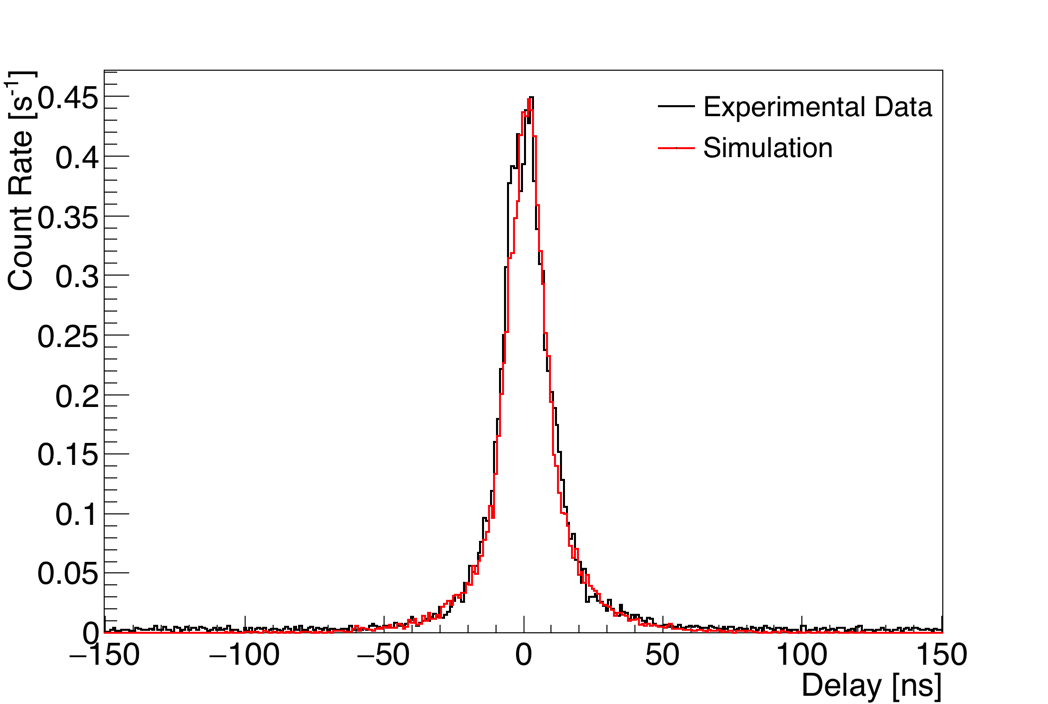} &
    \simsubfigimg[width=\linewidth]{\hspace{1.2cm}(b)}{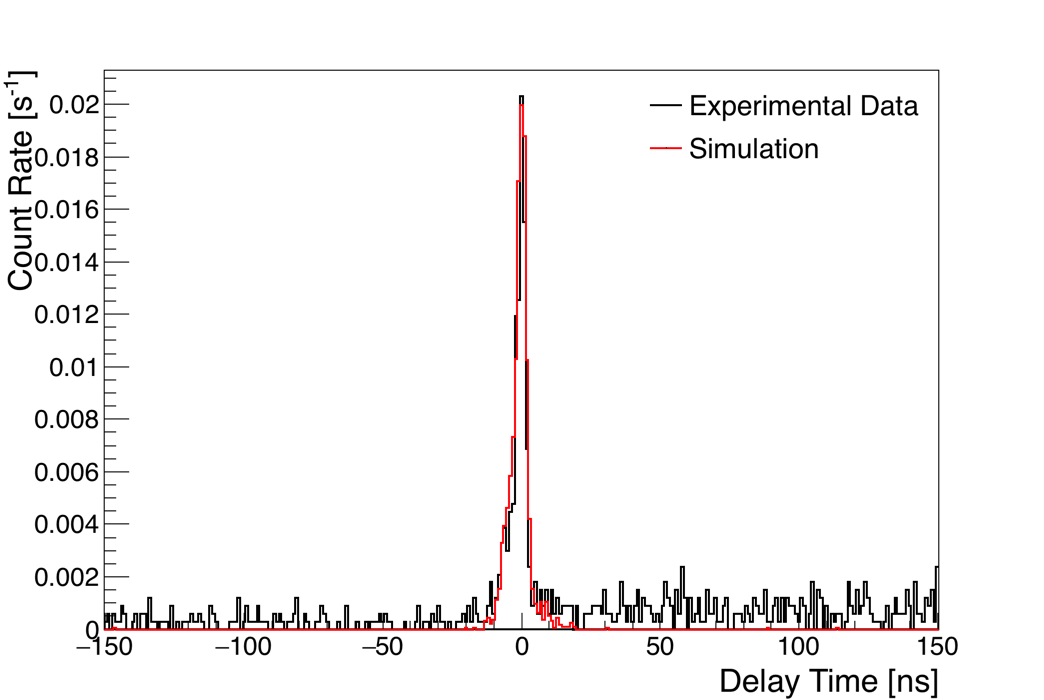} \\
  \end{tabular}
  \caption{Coincidence time distributions using the composite-EJ309 detector pairing for (a)~HEU and (b)~DU, where only neutron recoil events are accepted in the EJ309 detector. Simulated distributions are overlaid in red.}
  \label{fig:Coincidence_NGamma}
\end{figure}

Fig.~\ref{fig:Coincidence_NGamma} shows the coincidence time distributions for HEU and DU after interrogation with the neutron generator when only neutron recoil pulses are accepted from the EJ309 detector, with simulation results overlaid. Notably, the neutron-based coincidence time distribution for HEU is much broader than the corresponding distribution for DU and distributions where all event types were accepted. This is due to differences in the time of flight for different neutron energies, and perhaps to some extent by the fact that coincident pulses may be produced by radiation from different generations in the fission chain reaction. The broadening of the HEU coincidence distribution is also accurately reflected in the MCNPX model, which has a full-width at half maximum (FWHM) of 15.7~ns, compared to 16.1~ns for the experimental data. The DU distribution also agrees well with simulation, including the asymmetry of the coincidence peak, which is caused by neutron events in the EJ309 detector. For this simulation, the best agreement with experimental data is achieved when only gamma-ray recoils are considered in the composite detector. This provides further support for the conjecture that low-energy neutrons often do not register a recoil pulse in the composite detector. 

\subsubsection*{\centering{Supplementary References}}
\vspace{-20pt}
\bibliography{DAFBib2018,rev-tex-custom}